\documentclass[10pt,letterpaper]{article}
\usepackage[top=0.85in,left=2.75in,footskip=0.75in]{geometry}

\usepackage{amsmath,amssymb}

\usepackage{changepage}

\usepackage{textcomp,marvosym}

\usepackage{cite}

\usepackage{nameref,hyperref}

\usepackage[table]{xcolor}

\usepackage{array}

\newcolumntype{+}{!{\vrule width 2pt}}

\newlength\savedwidth

\raggedright
\usepackage[aboveskip=1pt,labelfont=bf,labelsep=period,justification=raggedright,singlelinecheck=off]{caption}

\makeatletter
\renewcommand{\@biblabel}[1]{\quad#1.}
\makeatother

\usepackage{lastpage,fancyhdr,graphicx}
\usepackage{epstopdf}
\fancyheadoffset[L]{2.25in}
\fancyfootoffset[L]{2.25in}
\begin{document}
\vspace*{0.2in}

\begin{flushleft}
{\Large
\textbf\newline{Random walk informed community detection reveals heterogeneities in the lymph node conduits network} 
}
\newline
\\
Solène Song\textsuperscript{1*},
Malek Senoussi\textsuperscript{1},
Paul Escande\textsuperscript{2},
Paul Villoutreix\textsuperscript{1*}

\bigskip
\textbf{1} Aix Marseille Univ, Université de Toulon, CNRS, LIS \\ Turing Centre for Living Systems, Marseille, France
\\
\textbf{2} Aix Marseille Univ, CNRS, Centrale Marseille, I2M, Marseille, France
\\
\bigskip

* paul.villoutreix@univ-amu.fr, solene.song@univ-amu.fr

\end{flushleft}
\section*{Abstract}
Random walks on networks are widely used to model stochastic processes such as search strategies, transportation problems or disease propagation. A prominent example of such process is the guiding of naive T cells by the lymph node conduits network. Here, we propose a general framework to find network heterogeneities, which we define as connectivity patterns that affect the random walk. We propose to characterize and measure these heterogeneities by detecting communities in a way that is interpretable in terms of random walk. Moreover, we use an approximation to accurately and efficiently compute these quantities on large networks. Finally, we propose an interactive data visualization platform to follow the dynamics of the random walks and their characteristics on our datasets, and a ready-to-use pipeline for other datasets upon download. 
By computing quantitative feature of random walk informed communities detected within the network, we show that the lymph node conduit network is spatially coherent,  however, despite its quasi-regularity, contains some random walk related heterogeneities. To evaluate these characteristics, we applied the same workflow of diffusion based community detection and analysis on the \textbf{LNCN} and a series of generated toy networks.

\section*{Author summary}
Lymph nodes are organs in which actors of the immune system meet. In particular, the encounter between the naive T cells and the antigens they are specific to occurs in lymph nodes. This event triggers the adaptive immune response. Lymph nodes are spanned by a conduits network, which was shown to serve as substrate for T cells migration, guiding them like railroads while they explore the lymph node looking for antigens. How does the connectivity pattern influence the exploration behavior of T cells, if they perform random walk on the conduits network ? Are there regions in the lymph node conduits network which have distinct random walk related properties ? The recently published topological reconstruction of the lymph node conduits network was recently made available. The network is very large (about 200 000 nodes) and appears very regular, with most nodes being connected to three neighbours. We propose a workflow to detect heterogeneity in such as large and quasi-regular network, which includes previously defined random walk on network tools, and the measure of two features which we interpret using a series of generated toy models for comparison. We show that the lymph node conduits network displays very localized small regions with distinct random walk properties. The rest of the network, which is most of its volume, promotes uniform exploration.


\section*{Introduction}
Random walks on networks are a widely used model to describe search strategies \cite{okubo1980diffusion,bartumeus2005animal}, transportation problems \cite{de2014navigability}, transmission in epidemiology \cite{lima2015disease,Alun_virus} or diffusion of information \cite{mislove2007measurement}. In this model, random walkers hop from node to node while choosing with an uniform probability the edges on which to travel. The structure of the underlying network, such as its degree distribution and connectivity pattern, will thus determine how the random walk evolves over time. Can the connectivity pattern favour the exploration of some nodes over others? We will refer to any feature of the connectivity pattern which can bias the exploration pattern of random walkers as 'local heterogeneities'. We aim at defining a measure of such heterogeneities that is applicable to large networks.  

Indeed, the question of the influence of network connectivity on random walker behavior is raised in the case of biological transportation networks. Among such networks, the lymph node conduits network (\textbf{LNCN}) offers a prominent example of a large network whose structure can affect its function \cite{kelch2019high}. The lymph nodes, among other functions, are hubs along the lymphatic system in which T cells encounter dendritic cells upon an infection. The dendritic cells bring the virus' antigen to the lymph node. Only a small subset of the naive T cells able to react to the antigen (one out of 1 000 000). When these relevant naive T cells encounter the dendritic cells, they proliferate and the specific immune response starts. This crucial encounter arises after a search phase \cite{krummel2016t}: The dendritic cells stay still at a certain location on the network, and the naive T cells scan the lymph node for dendritic cells to test their specificity. The \textbf{LNCN} is a network of pipes conveying lymph that span the lymph nodes. The naive T cells were shown to use the conduits as support for their migration \cite{bajenoff2006stromal}. Thus, without additional hypothesis, the modality of T cells search behavior is a random walk guided by the lymph node conduit network (\textbf{LNCN}). Does the network connectivity optimize the search by making some regions more accessible? 

Previous studies have shown that the network formed by the cells that cover the conduits, the FRC (fibloblastic reticular cells), has a small-world structure and shows high resilience to partial ablation \cite{Novkovic_conduits}. This network is surprisingly different from the network formed by the conduits themselves (\textbf{LNCN}) which exhibits less small-world characteristics and lower resilience. These characteristics were measured on a slice of a mouse lymph node. More recently, the whole conduits network (\textbf{LNCN}) was imaged by Kelch et al. \cite{kelch2019high}. The authors observed that there is a higher density of nodes on the periphery on the network and lower density at the core. Focusing on the guiding function of the conduits network (\textbf{LNCN}), the authors performed an agent-based model to simulate T cells migration which concluded that immune cells would have similar behaviors in both regions. Like them, we will consider the \textbf{LNCN} as the substrate of migration of the T cells, and not the FRC network, and thus base our analysis on their \textbf{LNCN} topological reconstruction. 

To answer the question of whether the network topology biases the search, measures such as the global mean first passage time (GMFPT)\cite{tejedor2009global} seem to be natural measures to detect nodes which are found by random walkers first. However this measure is local (node by node) and does not immediately distinguish regions at scale of interest, and it is not computable for large networks since it requires the computation of all return probabilities for all nodes at all time steps. Here we propose to assess the level of heterogeneity of the network in the sense of comparing the random walk related properties between regions, i.e. communities of nodes, of the network. If these features are similar in all regions, the network is homogeneous. On the contrary, if some regions features differ considerably from the others, the network displays heterogeneity. To devise random walk interpretable communities, we compute clusters in the so-called diffusion space \cite{pons2005computing}\cite{noh2004random}. These communities can be interpreted as groups of nodes which are highly connected by random walks paths, which means that there are many short paths connecting the nodes\cite{coifman2006diffusion}. Additionally, random walkers departing from nodes of a same community have correlated probability of presence fields over time. These communities can be defined at different time scale and are tractable for large networks.

In the present study, we compute these diffusion communities in the \textbf{LNCN}, and we introduce two measures on each community. These two measures, the Cheeger mixing index, and the mean entry and exit probabilities, aim at answering two different questions. The first question is : Do the diffusion communities form compact groups of neighbouring nodes or are they on the contrary scattered across the network ? The latter case implies that in some cases there is more chance to reach a node at other end of the network than a node which is only a few edges apart. We call this property "spatial coherence", which describes how the diffusion accessibility is correlated with the shortest paths lengths. This question is motivated by a seemingly contradiction in the description of the lymphatic network in the lymph node. Indeed, on one hand, the conduits network that was described qualitatively as a mesh, \cite{kelch2019high}, a loosely defined concept that suggests high spatial coherence. On the other hand, the small-world property of the FRC network \cite{FRC_small_world} suggests that there are shortcuts between otherwise distant nodes of the network \cite{watts1998collective}. To answer this question, we introduce the Cheeger mixing index. Using toy models, we show that the Cheeger mixing index is particularly low for planar networks. Then, we show that the Cheeger mixing values in the \textbf{LNCN} are as low as in 3D Voronoi tesselations of same sizes, which are a 3D analog of planar networks in 2D.

The second question is : is the network heterogeneous? This will be assessed by measuring the mean entry and exit probabilities of diffusion communities and assessing their variability.
To answer the second question, we measure the mean entry and exit probability at relaxation time, a reference time that can be found in all networks. We show that the mean entry and exit probabilities are able to distinguish two very similar quasi-regular toy networks, proving its sensitivity to heterogeneity beyond the effect of the degree distribution. Then, comparing the levels  of variation of the values between communities of the \textbf{LNCN} with null models of the same size, we conclude that the \textbf{LNCN} significantly heterogeneous. We locate a few communities with lower mean entry and exit probabilities at the time scale of the exploration time of the T cells. These communities are at the extremities of the longest axis and one near the medulla which is where the T cells exit the lymph node. The rest of the \textbf{LNCN}, which is most of its volume, appears homogeneous, promoting an uniform exploration by random walkers.

In summary, we exploit the interpretability of previously defined diffusion space coordinates \cite{pons2005computing}\cite{noh2004random}\cite{coifman2006diffusion}\cite{nadler2005diffusion} and its associated approximation \cite{nadler2005diffusion} to propose a workflow to detect heterogeneity in large networks. The value of the study is (i) to show the power of diffusion communities in detecting differences in connectivity patterns illustrated by toy models (ii) provide an overview of the \textbf{LNCN} structure in relation to T cells exploration behavior. Furthermore (iii) our analyses are available on the interactive visualization platform  and the workflow can be executed on any other dataset upon download.

\section*{Materials and methods}

In this section, we first describe our workflow, and the datasets studied in this paper. The workflow, which aims at detecting region-wise heterogeneities, consists in defining communities within the network which are interpretable from a random walk point of view, and to measure some features in each community to detect specific regions with outlying values for these features. As for the datasets, we introduce the \textbf{LNCN} which is the biological network of interest, and 8 other generated networks of different classes and sizes to help interpret the measured features.

\begin{figure}
\includegraphics[width = 0.4\textwidth]{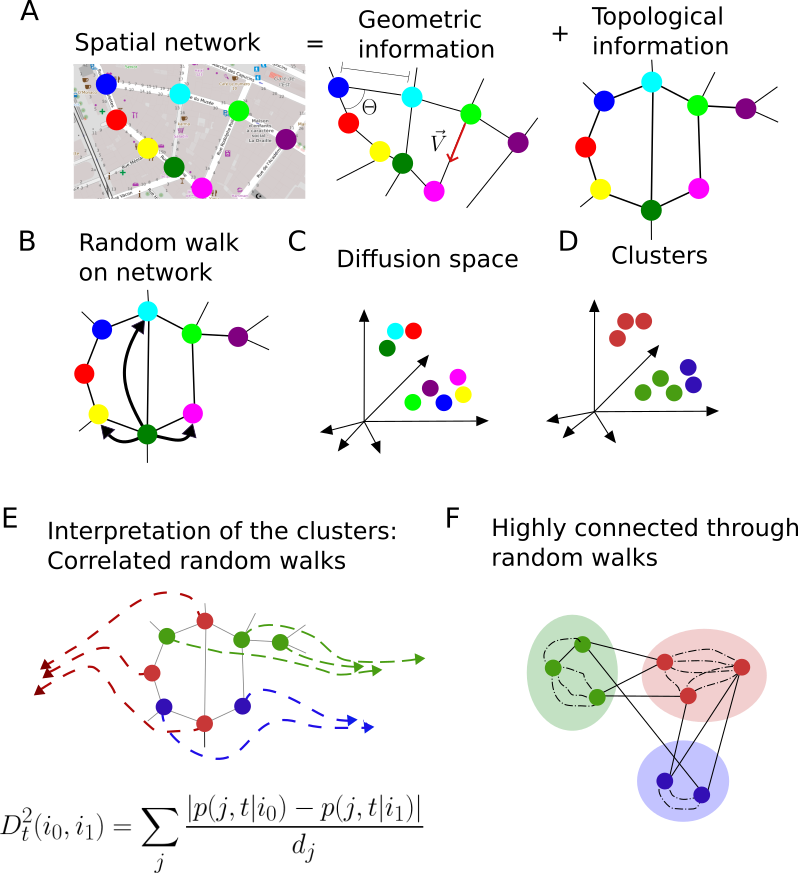}
\caption{Description of the diffusion communities detection workflow A: Decoupling of a spatial network into geometric and topological information B: We consider a random walker on the topological network C:Nodes of the network are embedded into the diffusion space D: Diffusion communities are clusters made in the diffusion space E: These communities are interpreted as: random walkers departing from nodes of same cluster follow correlated probability of presence fields F: Nodes are close in the diffusion space if they are highly connected through random walks. Dash lines represent random walks trajectories through nodes not represented on the sketch.}
\label{fig:concept}
\end{figure}

\subsection*{Workflow \label{met:workflow}}

We consider only the connectivity information of the network under consideration and ignore the spatial coordinates and edges lengths. Therefore, we address a discrete-time random walk on an unweighted, undirected network. The dynamics of a random walk on a network can be derived analytically . We use previously defined \textit{diffusion coordinates}, based on the definition of a random walk\cite{coifman2006diffusion,pons2005computing}, to compute community detection. The workflow is illustated by Figure \ref{fig:concept}.

\paragraph{Random walk on a network}
Let $\mathcal{N} = (\mathcal{V},\mathcal{E})$ be a spatial network where $\mathcal{V} \in \mathbb{R}^{3 \times N}$ is the set of nodes and $\mathcal{E} = \{ (i,j) \in [1,N]^{2}\}$ is the set of edges. The connectivity information is encoded in the adjacency matrix $A$ which is defined as $A_{ij} = 1$ if $(i,j) \in \mathcal{E}$ and $A_{ij} = 0$ otherwise. We consider a random walker following a Markovian process on the network. At each time step, the walker chooses with equal probability to jump to one of its adjacent nodes. The transition matrix $T=D^{-1} A$ (where $D=diag(d_{i})$ is the diagonal degree matrix) encodes the probabilities of transition from one node to another in one time step $T_{ij}=p(j,t=1|i)$.

\paragraph{Diffusion coordinates}
The diffusion coordinates allow to embed each node of the network in an interpretable euclidean space. These distances depend on the random walk on the network, thus the embedding is called the \textit{diffusion space}. \cite{nadler2005diffusion}. The diffusion coordinates are computed by using the spectral decomposition of the transition matrix T. T allows a spectral decomposition with left and right eigenvectors \cite{coifman2006diffusion,nadler2005diffusion,lee2019network} $T=\Psi\Lambda\Phi^{T}$, and for any time step $t=q$, $T^{q}=\Psi\Lambda^{q}\Phi^{T}$. The diffusion coordinates for a given time $t=q$ are given by the rows of the left part of the spectral decomposition: $X(q)=\Psi\Lambda^{q}$. The eigenvectors of the transition matrix were shown to be a discrete approximation of the eigenfunctions of a Fokker-Planck operator, which describes the evolution of brownian motion \cite{nadler2005diffusion}.
The euclidean distances in this embedding, also called diffusion distances $ D^2_t(i_0,i_1) =\lVert \Psi_t(i_0) - \Psi_t(i_1) \rVert^2 =\lVert p(j,t \vert i_0) - p(j,t \vert i_1) \rVert^2_w$ 
are interpreted as a measure of the correlation of random walks departing from $i_0$ and $i_1$, with $(i_0,i_1)\in [1,N]^2$ : 
\begin{align}
    D^2_t(i_0,i_1) &= \lVert p(j,t \vert i_0) - p(j,t \vert i_1) \rVert^2_w \\
    &= \sum_{y}{(p(j,t \vert i_0) - p(j,t \vert i_1))^2 w(j)} 
\end{align} where $w(j) = 1/\phi_{j0}$, with $\phi_{j0}=\frac{d_{j}}{\sum_{i}d_{i}}$ \cite{nadler2005diffusion}.
Furthermore, nodes are close in the diffusion space if random walkers are likely to travel from one to the other through short paths. Depending on the network, this distance can differ considerably from the topological distance in the network, which is the shortest path length between nodes.   
For large networks, the spectral decomposition of the transition matrix is computationally impossible. We therefore use an approximation.

\paragraph{Approximation}
For large number of nodes, such as for the \textbf{LNCN}, which contains approximately 200,000 nodes, the computation of $\Psi$ and $\Phi$ requires the diagonalization of the symmetrized transition matrix $T_{s}$, which cannot be completed fully for large $n$. The matrix $T$ and the diffusion coordinates $X(q)$ can be respectively approximated as the truncated spectral decomposition $\hat{T}^t_K = \Psi_K\Lambda_K^t\Phi_K^{T}$ where $\Psi_K = \left(\psi_0,..,\psi_K \right)$, $\Phi_K = \left(\phi_0,..,\phi_K \right)$ and $\Lambda_K = \text{diag}\left( \lambda_0, ... , \lambda_K\right)$, and $ D^2_t(i_0,i_1) = \sum_{k = 1}^{K}{\lambda_k^{2t}. \left(\psi_k(i_0) - \psi_k(i_1) \right)^2}$, as detailed in \nameref{S2_Appendix}.

Because $1=|\lambda_{0}|>|\lambda_{1}|\geq...\geq|\lambda_{N-1}|\geq0$, the larger $t$ is, the more negligible the last terms (large $j$) are.
To compare the actual $T^t$ and its approximation $\hat{T}^t_K$, we considered the spectral norm of their difference (i.e. $\|T^t - \hat{T}^t_K\|_{2,2}$) which can be computed with the power iteration method \cite{golub1996matrix}. The computation is described in more details in \nameref{S3_Appendix} for more details. The relative error $\frac{\|T^t - \hat{T}^t_K\|_{2,2}}{\|T^t\|}$  falls below $1\%$ for t larger than 250, and below $0.1\%$ for t larger than 350 as shown in \nameref{S1_Fig}. In this paper, the time scale we consider are even larger, superior to t=900, so that the accuracy of the approximation is guaranteed. 

\paragraph{Community detection and relaxation time}
Using the diffusion coordinates, the nodes of the network can be seen as a point cloud. To identify subset of nodes having similar properties, we computed communities using the k-means algorithm in the diffusion coordinate space. Since the diffusion coordinate are dependent of the time $t$, we chose to always use a reference time in the dynamics of the random walk, $\tau$ the relaxation time. The relaxation time is the time at which the difference between the probability field and the stationary field is reduced significantly by a constant factor. The relaxation time differs from one network to the other and provides a comparable time scale from one network to another. This time is governed by the ratio between the magnitude of the second largest eigenvalue of the transition matrix $|\lambda_{1}|$ and $\lambda_{0}=1$, $\tau=\frac{1}{1-\lambda_{1}}$. The derivation is in \nameref{S4_Appendix} and the values for all the considered networks are summarized in Table \ref{tab:networks_char}. As for $k$, the number of clusters, we chose to stick with an arbitrary $k=100$ for all the large networks of similar sizes as the \textbf{LNCN} and $k=4$ for the smaller networks of 2050 nodes. We fixed $k$ to be the same for all networks of the same size because the Cheeger mixing value that we present below is highly dependent on the size of the communities and hence the number of communities, as illustrated by \nameref{S4_Fig}.

\vspace{3 pt}

To analyze and understand the properties of the communities, we compute two features, the Cheeger mixing, and the mean entry and exit probabilities. The Cheeger mixing measures to what extent the nodes which belong to the same diffusion community, thus which are close in the diffusion space, form compact neighbourhoods in the network. The mean entry and exit probabilities are features that can be computed over time, and can be compared between communities to distinguish outlying communities with especially low or high values. 

\begin{description}
\item[Cheeger mixing] The Cheeger mixing for the community C
$h(C)=\frac{\sum_{i \in C}\sum_{j \in \Bar{C}} A_{ij}}{min(\sum_{i \in C}d_{i}, \sum_{i \in \Bar{C}} d_{i}}$
measures the relative number of edges connecting a node of C and a node that does not belong to C. If it is large it means that the nodes that belong to the same diffusion community are scattered across the network. We call this value Cheeger mixing because the minimal value over all possible sets of nodes instead of community C is known as the Cheeger constant. 
\item[Mean entry and exit probabilities] 
For each cluster, the mean entry and exit probabilities were computed respectively as $<p_{in}>_C(t)=\frac{1}{n_{C}n_{\overline{C}}}\sum_{l\in C}\sum_{m\in \overline{C}}\sum_{j=1}^{k}\psi_{j}(m)\lambda_{j}^{t}\phi_{j}(l)$
and $<p_{out}>_C(t)=\frac{1}{n_{C}n_{\overline{C}}}\sum_{l\in \overline{C}}\sum_{m\in C}\sum_{j=1}^{k}\psi_{j}(m)\lambda_{j}^{t}\phi_{j}(l)$. Note that (i) if a network is fully regular, $<p_{in}>_C(t)=<p_{out}>_C(t)$ for all C and (ii) $<p_{in}>_C(t)\underset{t\to +\infty}{\longrightarrow} \frac{1}{d_{tot}}\frac{\sum_{i \in C}d_i}{|C|}$ and $<p_{out}>_C(t)\underset{t\to +\infty}{\longrightarrow} \frac{1}{d_{tot}}\frac{\sum_{i \in \Bar{C}}d_i}{|\Bar{C}|}$, where $d_{tot}=\sum_{i=1}^{N} d_{i}$ (derivation in \nameref{S9_Appendix}). At long time scale, these measures depend only on the degree distribution among the communities but their computation at shorter time scales prove useful to distinguish communities, especially when the networks are quasi-regular as shown in Results section.
\end{description}

\subsection*{\label{sub_sec:networks} Datasets \protect}

\paragraph{The lymph node conduit network (\textbf{LNCN})} The network's connectivity, published in \cite{kelch2019high}, was extracted from a segmented 3D microscopy acquisition of a whole mouse popliteal lymph node conduits network (850 x 750 x 900 $\mu$m) obtained from microscopy data \cite{kelch2019high}in which the conduits are made fluorescent by injection of labelled molecular tracer into the lymphatic vessels. The conduit network is restricted to the T zone in which T-cells are present. This is a 3D network made of 192,386 nodes and 274,906 edges.  Most of its nodes are degree 3 ($72\%$) , and $99\%$ of nodes having degree between 1 and 4 (see degree distribution in Figure \ref{fig:LN}.A). 

\paragraph{Homogeneous City Network (\textbf{HCN}) and Polar City Network(\textbf{PCN})}

These two networks \cite{pousse2020caracterisation} (details on the model can be found in \nameref{S5_Appendix}) have both the same number of nodes N=2050, number of edges E=3073 and degree distribution (degree 3 for all nodes except from the 4 nodes on the corners) thus they are highly regular and very similar to each other. These networks were initially  spatial: the nodes were associated to 2D coordinates and the edges links them into rectangular cells, as shown on their upper panel of Figure \ref{fig:HCN_PCN}. For the \textbf{HCN}, the density appears uniform whereas for the \textbf{PCN}, there are two high density regions, near the left and right borders. On the bottom panels, the same networks are represented using only the connectivity information : which nodes are connected to which and which ones are not. The layout is obtained using a force-directed drawing algorithm \cite{fruchterman1991graph}, which minimizes the occurrence of edges crossing and edges as equally long as possible. 

\paragraph{Geometric network}
The geometric network was obtained using the network generator from Python library NetworkX \cite{hagberg2008exploring} random\_geometric\_graph \cite{penrose2003random}. N=2050 nodes positions were drawn from a set of nodes which locations x and y are drawn from Gaussian distribution centered on (0,0) and with standard deviation 2. All the nodes which are closer than the chosen threshold 2.1 are connected by an edge. There are 479,786 edges. 

\paragraph{Regular random network (\textbf{RRN})}
Regular random networks are networks in which the nodes are randomly connected, with the constraint that all nodes should have a given defined degree. The regular random networks were generated using the network generator from Python library NetworkX \cite{hagberg2008exploring}  random\_regular\_graph \cite{steger1999generating}\cite{kim2003generating}. We generated two regular networkss of degree 3 : one small network with 2050 nodes and 3075 edges, one big network with 192,386 nodes and 288,579 edges.

\paragraph{Erd\H{o}s-R\'enyi \textbf{ER}}
Erd\H{o}s-R\'enyi networks are random networks in which the presence of an edge between two nodes follows a fixed probability $p$, independent from the other edges.
The Erd\H{o}s-R\'enyi networks was generated using the network generator from Python library NetworkX \cite{hagberg2008exploring} erdos\_renyi\_graph \cite{erdHos1959random} \cite{gilbert1959random} with N=2050 and p=0.005. It has 42,009 edges.

\paragraph{Homogeneous Voronoi 3D}
The homogeneous Voronoi 3D was generated using the network generator from Python library Scipy \cite{2020SciPy-NMeth} spatial.Voronoi. The input points coordinates were defined such as : (i) create a rectangular 3D grid of 27x27x27 nodes (ii) keep only nodes inside a sphere resulting in 28991 nodes (iii) introduce noise in the coordinates so that the grid structure is not perfect. Indeed, the case in which the 3D grid is perfect corresponds to a limit case where Voronoi vertices are degree 6 instead of 4. The resulting Voronoi diagram contains 197,123 nodes and 393,527 edges.

\paragraph{Polar Voronoi 3D}
The polar Voronoi 3D was generated using the network generator from Python library Scipy \cite{2020SciPy-NMeth} spatial.Voronoi. The input points coordinates were defined such as: the union of two gaussian distributions in 3D, one centered in (0,0,0) with standard deviation 5, with $N_1=10,000$ nodes, and the other one centered in (5,5,5) with standard deviation 3 with $N_2=18,752$. The resulting network contains 193,391 nodes and 386,734 edges.

\subsection*{\label{sub_sec:int_viz}Interactive visualization tool}

The method developed in this paper enables to monitor the probability density of a random walk from any starting point after any number of time steps $p(j,t \vert i)$. As tracking the density of probability distributions over time is useful to explore visually the topology of a network, we developed an interactive visualization platform available at this address: \url{https://randomwalknet.centuri-engineering.univ-amu.fr/}. The visualization platform is divided into three parts. The first part contains 2 animations, one is the probability of presence field of a random walk starting from a given point, and the other is the probability field integrating on all departure nodes. The second part contains the analyses of the networks based on classical indicators: betweenness centrality, closeness centrality, eigenvector centrality, as described in \nameref{S6_Appendix}.
The third part shows the results of the random-walk informed community detection algorithm.

\section*{Results}

\begin{table}[b]
\caption{\label{tab:networks_char}%
Summary of the characteristics the considered networks } \label{tab:relax}
\begin{tabular}{cccc}
Model & Size & Relaxation time & Regularity \\
\hline
\textbf{HCN} & 2050 & 1111 & almost all nodes degree 3   \\
\textbf{PCN} & 2050 & 1759  & almost all nodes degree 3 \\
\textbf{Geometric network}& 2050 & 4 & no\\
\textbf{small RRN} & 2050& 17 & all nodes degree 3\\
\textbf{ER} &2050& 1 & no\\
\textbf{HVor}&197123 & 2854 & almost all nodes degree 4\\
\textbf{PVor}& 193391 & 2941 & almost all nodes degree 4 \\
\textbf{large RRN} & 192386 & 17 & all nodes degree 4\\
\textbf{LNCN}& 192386 & 17797& most nodes degree 3\\
\end{tabular}
\end{table}

\subsection*{\label{sub_sec:LN} Despite quasi-regularity, the \textbf{LNCN} displays random walk related heterogeneities}

The \textbf{LNCN}, for which the topology was reconstructed by Kelch et al. \cite{kelch2019high}, is a large network in which a large proportion of nodes have the same degree, which is 3. Its size restricts the measures that can be made on the network because of the computational challenges, and its quasi-regularity gives a first impression of uniformity. Thus, to address the detection of heterogeneity in the network, in the sense of detecting regions of the network that are distinct from the others with respect to the process of random walk on the network, we apply the workflow described above in the Methods section.

\begin{figure*}[!ht]
\includegraphics[width = 0.8\textwidth]{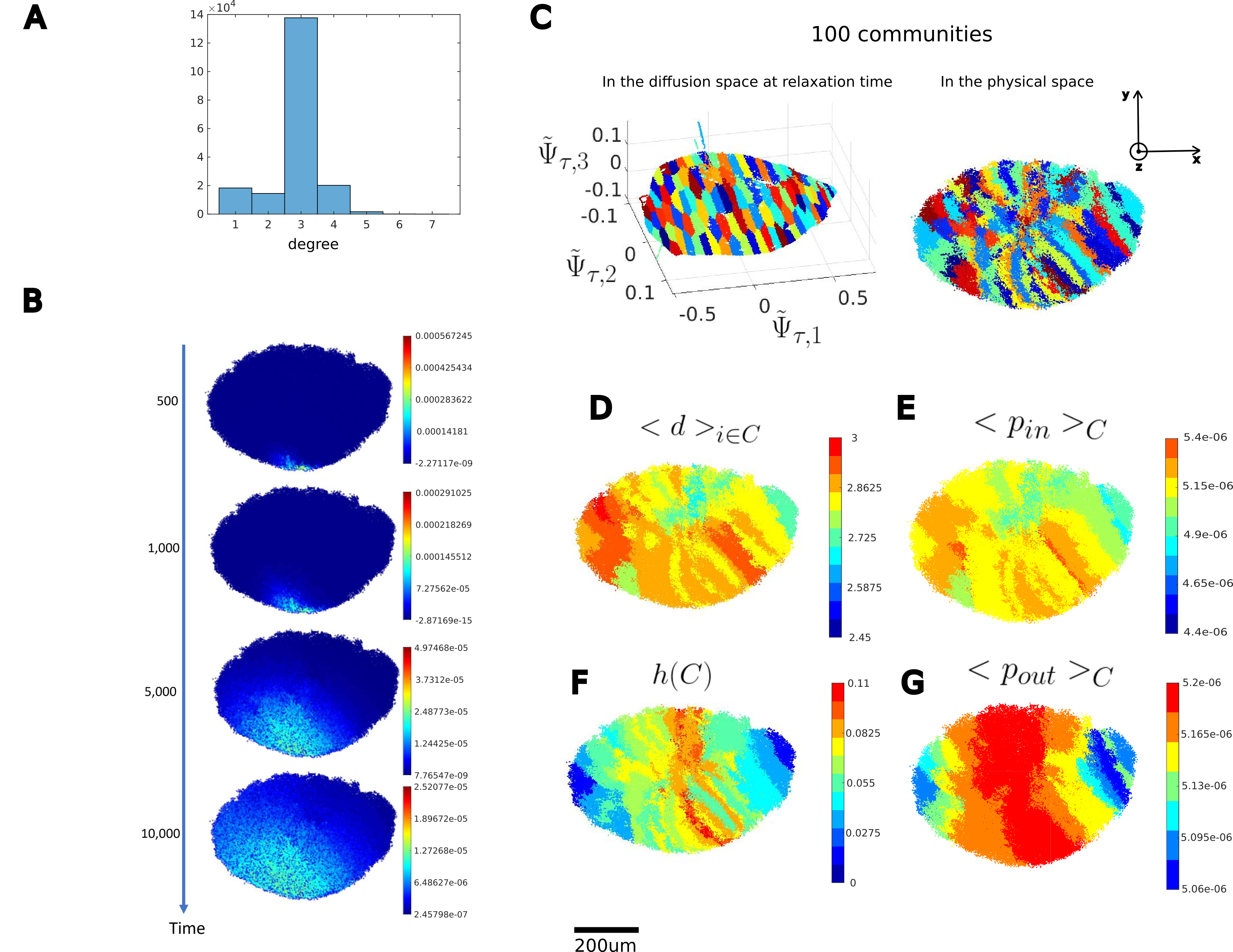}
\caption{\textbf{LNCN} features. A: Degree distribution B: Probability of presence departing from one chosen node in time C 100 diffusion communities shown in the diffusion space 3 first coordinates (left panel) and the physical space (right panel) D: mean degree for each community for relaxation time $\tau$ E: mean entry probability for each community F: Cheeger mixing for each community  $\tau$ G: mean exit probability for each community for relaxation time}
\label{fig:LN}
\end{figure*}

\subsubsection*{Tracking random walk in the \textbf{LNCN} highlights the spatial organization of the network}

For any time t, the probability of presence of the random walker at time t departing from a given node can be read as a term of the transition matrix elevated at the power t. However, the elevation at power t of the transition matrix becomes intractable when t grows higher and the matrix loses its sparcity. Thus, we apply the approximation described in the Methods section to have an approximation of the transition matrix.
We can visualize and follow over time at low computational cost the field of probability of presence of a random walker over time departing from a given node, as shown on the online tool introduced in the Methods section and in \nameref{S6_Appendix}. Figure \ref{fig:LN} shows these fields departing from a given node from time 500 to 10 000 (for comparison, the relaxation time is reached at  $\tau=17797$). The chosen node is the one with smallest y value. It is located at the center of the x axis (longest axis). We see that the probability field in time is skewed towards the left part on the x-axis. Other probability fields with other departure nodes can be seen with the online tool. The anisotropy of these fields hints towards a non uniform spatial organization of the network. Furthermore, the probability fields provide a way to explore the trajectories of random walkers departing from specific places chosen according to biological hypotheses without making an agent-based model.

\subsubsection*{Our analysis reveals a relationship between the communities in the diffusion space and in the physical space \label{space_coherence}}

100 diffusion communities were defined on the \textbf{LNCN} as described in the Methods section by using approximated diffusion coordinates at $t=\tau=17797$. These diffusion communities are shown on Figure \ref{fig:LN}C. The cluster size varies from 272 to 4905 nodes and has an average value of 1924 nodes.

Before assessing the question of the level of heterogeneity of the network by comparing features between the different communities, let us draw a few general characteristics of the network. We first note that the distribution of points in the diffusion space is essentially 2D as shown on Figure \ref{fig:LN}C and that the relative positions of the communities look the same in the diffusion space and the physical space. The shapes of the communities in the physical space are elongated, except at the two x ends (along the longest axis). The nodes belonging to the same communities seem to be close in the physical network. The average Cheeger mixing value $\bar h$ is 0.069, which we assume is low. This will be confirmed in a following section by comparing with similar sized networks. Low value of average Cheeger mixing indicates overall spatial coherence, as illustrated in a following section.

\subsubsection*{The communities display different values of mean entry and exit probabilities $<p_{in}>_C$ and $<p_{out}>_C$ and Cheeger mixing $h(C)$}

Overall the network is quite regular (more that $72\%$ of nodes of degree 3) although there is a small amount of nodes of higher degree. Despite this first impression of uniformity, measuring features defined in the workflow over each community highlights variations across communities. The measured features are the mean degree, the mean entry and exit probabilities, and the Cheeger mixing for each community. We expose the values in this section and we will make sense of the actual values in the next sections by comparing to generated toy models of known classes.

\begin{itemize}
    \item We measured the mean degree in each community to check if there are specific communities which are enriched for nodes whose degree is higher (or lower) than 3. We show the colormap in Figure \ref{fig:LN}. There are slightly higher degrees in the bottom half and left end. The mean degree peaks at 2.8 and varies between the minimal value of 2.45 and the maximal value of 3, with standard deviation 0.06. For comparison, when the degrees of $N_c$ nodes are randomly sampled 1000 times from the empirical degree distribution, where $N_c=1924$ is the average size of diffusion communities in the \textbf{LNCN}), the mean degree follows a distribution that peaks at 2.85, with standard deviation 0.02, minimum value 2.81 and maximum value 2.91. Thus, the distribution of degrees across the communities span a larger range of variation than expected by drawing randomly from the empirical degree distribution.
    \item We compute the mean entry and exit probabilities $<p_{in}>_{C}(\tau)$ and$<p_{out}>_{C}(\tau)$ for each community $C$. The entry probability $<p_{in}>_C$ colormap shows some similarity to the one of degree colormap, which is expected at sufficiently long time scales since $<p_{in}>_{C}(t)\underset{t\to +\infty}{\longrightarrow} \frac{1}{d_{tot}}\frac{\sum_{i \in C}d_i}{|C|}$, as shown in \nameref{S9_Appendix}. The $<p_{out}>_C$ displays higher values in the center and lower at the extremities along the x axis.
    \item  The Cheeger mixing $h(C)$ also varies along the x-axis, with higher values in the central bands and lower values on the extremities.
\end{itemize}

The following sections aim at interpreting the results obtained on the \textbf{LNCN}. How low is the average Cheeger mixing value compared to other networks of different known classes?
How significant are the amplitude of variation of the mean entry and exit probabilities between communities? What kind of heterogeneity are these variations able to capture?

\subsection*{$<p_{in}>_C$ and $<p_{out}>_C$ measures are able to detect heterogeneities within quasi regular toy networks \label{pin_pout_cities}}

\begin{figure*}[!ht]
\centering
\includegraphics[width = 0.5\textwidth]{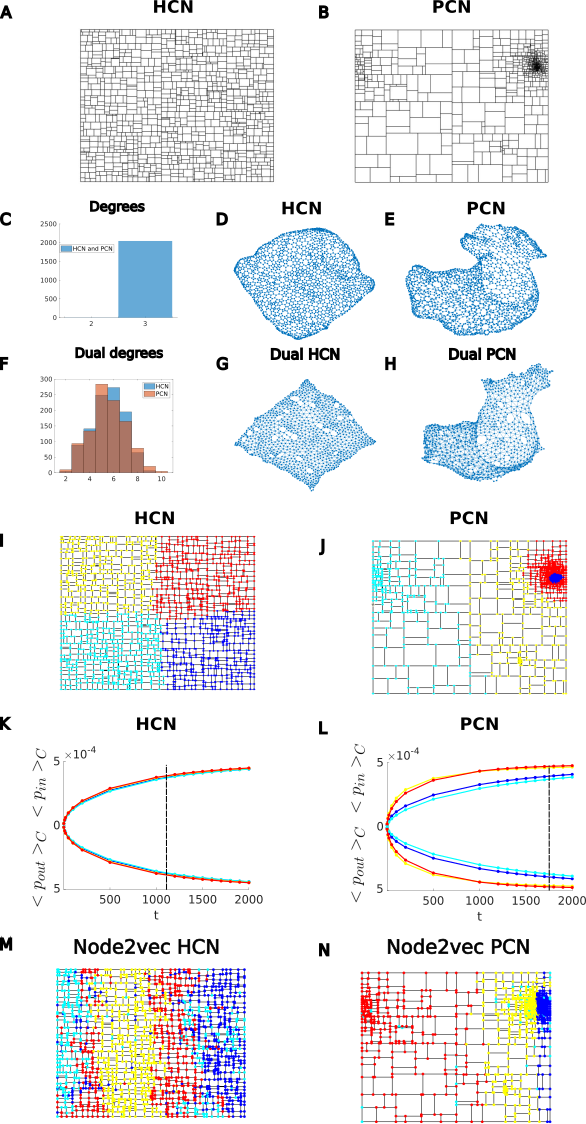}
\caption{The measure of$<p_{in}>_C$ and $<p_{out}>_C$ for the diffusion communities allows to distinguish between two very similar networks, the \textbf{HCN} and \textbf{PCN}. A,B:the networks \textbf{HCN} and \textbf{PNC} in their original spatial embedding C:degree distributions of the primal networks  D,E: representation of \textbf{HCN} and \textbf{PCN} using only connectivity information with force-directed drawing algorithm \cite{fruchterman1991graph} which minimizes the occurrence of edges crossing and edges as equally long as possible, F: degree distributions of the dual graphs G,H: dual network force layout representation of \textbf{HCN} and \textbf{PNC} I:4 diffusion communities in the \textbf{HCN} J:4 diffusion communities in the \textbf{PCN} K:$<p_{in}>_C$ and $<p_{out}>_C$ for different times for the diffusion communities in \textbf{HCN} K:same for \textbf{PCN} M,N:Node2Vec communities in \textbf{HCN} and \textbf{PCN}}
\label{fig:HCN_PCN}
\end{figure*}

To evaluate the power of $<p_{in}>_C$ and $<p_{out}>_C$ to highlight differences between communities, we first focused on the \textbf{HCN} and \textbf{PCN} networks, as shown on figure \ref{fig:HCN_PCN}.

The \textbf{HCN} and \textbf{PCN} are generated as spatial networks (the nodes are provided with spatial coordinates in 2D) \cite{pousse2020caracterisation}. Details on the model can be found in \nameref{S5_Appendix}. Their interest for testing our workflow is that they are extreme cases of quasi-regular networks, very similar to one another. They both have the same number of nodes N=2050, number of edges E=3073 and degree distribution (degree 3 for all nodes except from the 4 nodes on the corners), and same clustering coefficient (4-clustering coefficient, which means that they have the same number of rectangular faces). Thus, they are very similar to each other and highly regular. On the upper panels of Figure \ref{fig:HCN_PCN} the nodes are represented with their 2D spatial coordinates as initially generated by the model. Despite their commonalities, they do not have the same connectivity, as illustrated by the force layout representation of their connectivity, and shown by their distinct dual networks degree distributions. If the $<p_{in}>_C$ and $<p_{out}>_C$ measures are able to identify differences between \textbf{HCN} and \textbf{PCN} it would be an indication of its discriminative power.

We show that the spatial organizations of the diffusion communities are different between \textbf{HCN} and \textbf{PCN}. We define communities in the \textbf{HCN} and \textbf{PCN} by k-means clustering in the diffusion space. The time $t$ chosen to compute the diffusion coordinates is the relaxation time $\tau$. The number of clusters was chosen to be 4 for both models, \textbf{HCN} and \textbf{PCN}. In \textbf{HCN} the 4 communities make a partition into equal quadrants, whereas in \textbf{PCN} the communities make concentric shapes with one community (in blue) surrounded by another one (in red).

Comparing $<p_{in}>_C$ and $<p_{out}>_C$ in diffusion communities allows to distinguish between the \textbf{HCN} and \textbf{PCN}. We compute the values of $<p_{in}>_C(t)$ and $<p_{out}>_C(t)$ for different times, shown in Figure \ref{fig:HCN_PCN}, with the relaxation time shown for reference with dashed lines. $<p_{in}>_C(t)$ and $<p_{out}>_C(t)$ are nearly identical for all times for \textbf{HCN}. On the contrary, in the \textbf{PCN} there are two communities with larger $<p_{in}>_C(t)$ and $<p_{out}>_C(t)$ (light blue and dark blue), and two communities with lower $<p_{in}>_C(t)$ and $<p_{out}>_C$ (red and yellow). As a side note, the communities with larger values seem to correspond to high density regions in the original spatial coordinates space.

Overall, we show that the characteristics of the clusters obtained in the diffusion space are successful at differentiating the \textbf{HCN} and the \textbf{PCN}.  Indeed the \textbf{HCN} appears homogeneous, since all the communities have the same  $<p_{in}>_C(t)$ and $<p_{out}>_C(t)$ whereas the \textbf{PCN} appears heterogeneous, since there are two communities among four which have lower $<p_{in}>_C(t)$ and $<p_{out}>_C(t)$). As a side note, the computation of $<p_{in}>_C(t)$ and $<p_{out}>_C(t)$ for the remaining three small toy networks are shown in \nameref{S3_Fig}

Additionally we used these two toy networks to compare the diffusion communities with another state of the art diffusion based communities detection methods, the Node2vec algorithm \cite{grover2016node2vec}, described in more details in \nameref{S7_Appendix}. Node2vec output is an embedding in which we can make communities in a similar way as in the diffusion space. Optimized communities all show a band-like structure, yielding a partition very different from the diffusion communities, as shown in \nameref{S2_Fig}. The optimization of the Node2vec parameters is described in \nameref{S8_Appendix}. 

\subsection*{The Cheeger mixing index measures the spatial coherence of networks \label{Cheeger}}

\begin{figure*}[!ht]
\centering
\includegraphics[width = 0.5\textwidth]{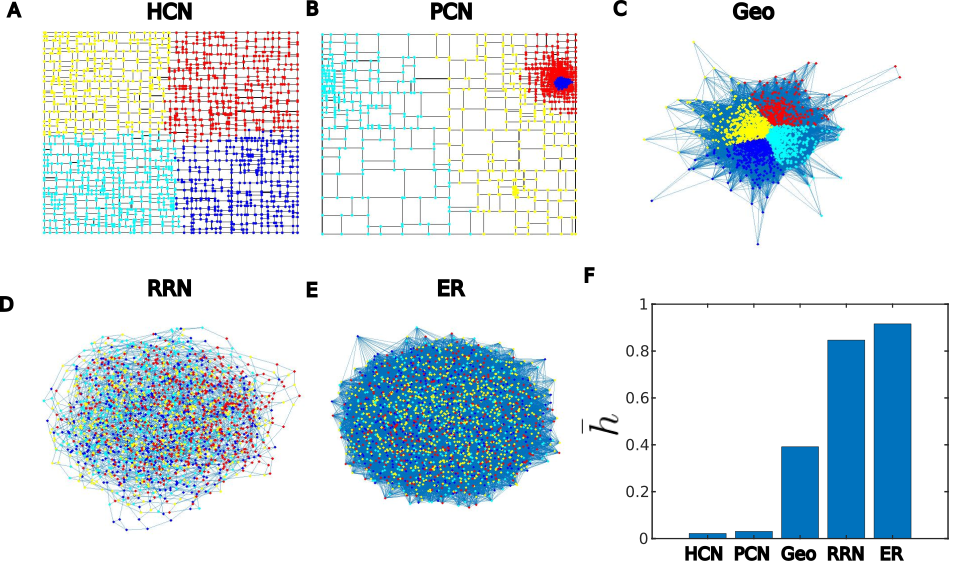}
\caption{Measure of  $\bar h$ on toy model shows that the values are low for networks in which there exists a spatial embedding where the existence of an edge is highly related to the distance between nodes in this embedding. A-E 4 diffusion communities for the \textbf{HCN}, \textbf{PCN}, \textbf{geometric network}, \textbf{small RRN}, and \textbf{ER}, F: $\bar h$ for each toy model. }
\label{mixing}
\end{figure*}

The Cheeger mixing measured on the diffusion communities, like introduced in the Methods section quantifies to what extent the nodes which are close in the diffusion space are also close neighbours in terms of shortest paths in the network. We call this feature spatial coherence. This measure is aimed at clarifying the structure of the \textbf{LNCN}. Indeed, as described in the introduction, the fact that the \textbf{LNCN} is called a mesh \cite{kelch2019high} suggests that we should find high spatial coherence. However, the FRC network that ensheath the conduits where shown to display small-world property \cite{FRC_small_world}. The small-world property implies that the network minimizes the mean shortest path length between two nodes and maximizes the clustering coefficient. As a consequence, there should be shortcut edges connecting distant nodes, thus low spatial coherence. 

The Cheeger mixing, measured on diffusion communities, can be interpreted as follows.
The diffusion distance $D_t^{2}(i_0,i_1)$ between two nodes $i_0$ and $i_1$ expresses the distance between the two posterior distributions $p(j,t|i_0)$ and $p(j,t|i_1)$. If $i_0$ and $i_1$ belong to the same community (i.e. are close in terms of diffusion distance), random walkers starting from $i_0$ and $i_1$ have correlated walks. Additionally,  $D_t^{2}(i_0,i_1)$ is small if there is a large number of short paths connecting $i_0$ and $i_1$ \cite{coifman2006diffusion}. Thus, a community defines a set of nodes that are highly connected and from which random walks are correlated. The Cheeger mixing of one community $C$ measures the proportion of edges between nodes of $C$ and nodes that do not belong to $C$ over the number of edges inside $C$. If this number is high we expect the nodes belonging to a same community to be scattered across the networks: close neighbours (in terms of shortest path length) are not necessarily the most well connected at this time t integrating on all possible paths. On the contrary if this number is low, the nodes belonging to the same community form a compact neighbourhood within the network, which means high spatial coherence.

To illustrate how the Cheeger mixing index captures the spatial coherence of networks, we compute it on different classes of toy networks. 
We compute the Cheeger mixing for five networks of same size (N=2050), as shown in Figure \ref{mixing}, partitioned into 4 diffusion communities. There are two planar networks \textbf{HCN} and \textbf{PCN}, and two random networks, \textbf{RRN} and \textbf{ER}. Like expected, the planar networks have very low Cheeger mixing values (averaged over all the 4 communities) since there are no paths connecting nodes that are far apart topologically. On the contrary, the random networks have high Cheeger mixing values. The last instance of toy network is a \textbf{geometric network}. The nodes are initially attributed space coordinates, and the nodes that are closer than a certain threshold get connected by an edge. This network has an intermediary mean Cheeger mixing value, higher than the planar networks but lower than the random ones. 
Therefore, we show that the Cheeger mixing measure is low when there exists a spatial embedding in which the probability to have an edge between two nodes is tightly dependant on the distance between the two nodes. This seems to be a consistent definition for quantifying the spatial coherence of a network.

We now have an interpretation of the $\bar h$ mean Cheeger mixing and the $<p_{in}>_C$ and $<p_{out}>_C$ measures. 
But these toy models are small. To interpret the values obtained on the \textbf{LNCN}, we would like to compare its features to similar size networks.

\subsection*{The \textbf{LNCN} is spatially coherent and significantly heterogeneous, compared to null models \label{LN_null}}

\begin{figure*}[!ht]
\centering
\includegraphics[width = 0.5\textwidth]{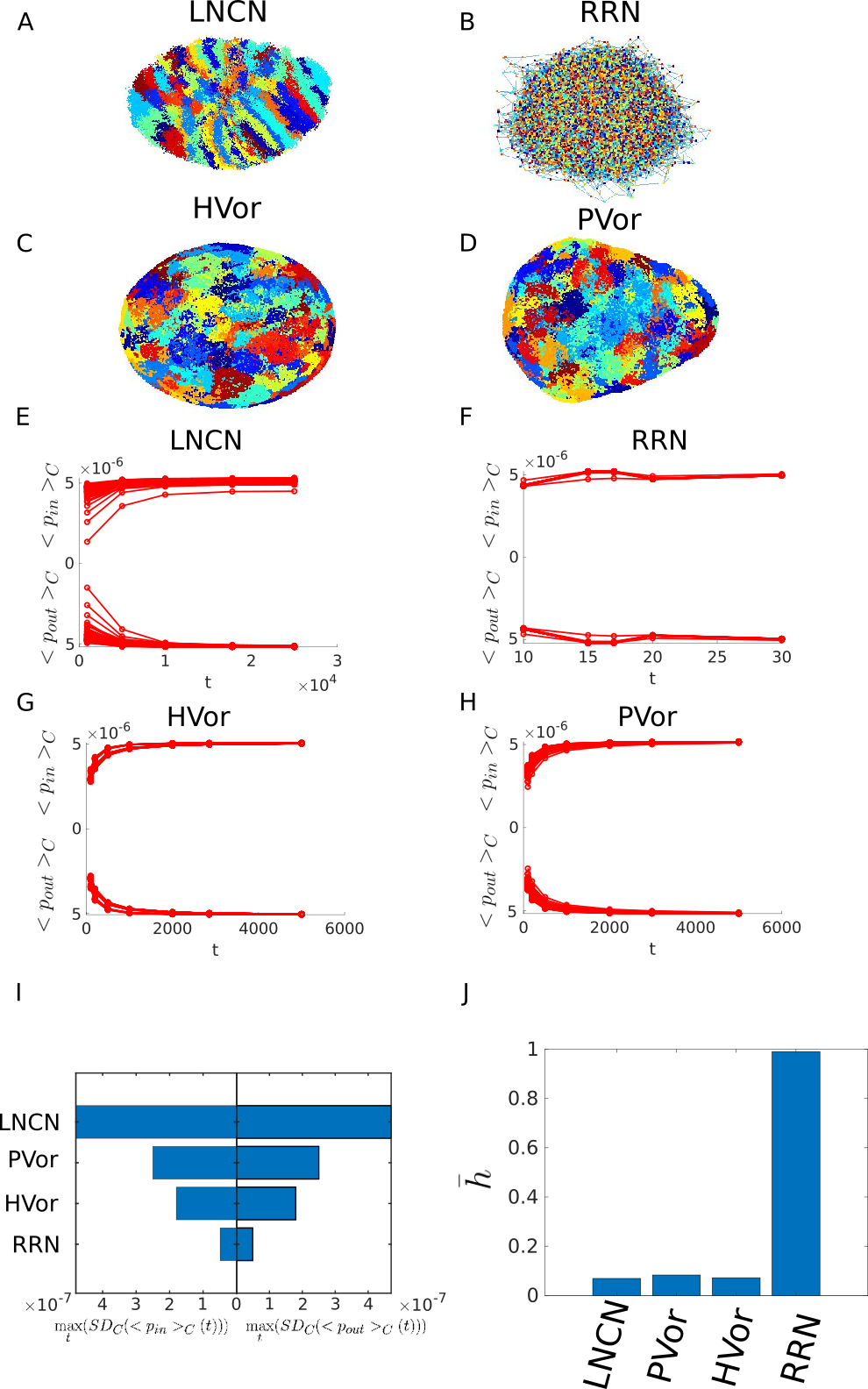}
\caption{The \textbf{LNCN} is as space-dependent as the Voronoi tesselations \textbf{HVor} and \textbf{PVor} and more heterogeneous than all 3 models, Voronoi tesselations and \textbf{RRN} A-D 100 diffusion communities in the \textbf{large RRN}, \textbf{HVor} and \textbf{PVor}. E-H: $<p_{in}>_C$ and $<p_{out}>_C$ in time for all communities for the networks \textbf{LNCN} \textbf{large RRN}, \textbf{HVor} and \textbf{PVor}. I:the maximal standard deviation  $<p_{in}>_C$ and $<p_{out}>_C$ across the 100 communities in each network, for all four large networks. J: the mean Cheeger mixing $\bar h$ for all four large networks}
\label{fig:big_graphs}
\end{figure*}

We generated three toy models of about the same size as the \textbf{LNCN}: \textbf{HVor} and \textbf{PVor} which are quasi-regular 3D Voronoi tesselations and \textbf{RRN} a random regular network. 

The \textbf{LNCN} is as spatially coherent as 3D Voronoi tesselations.
By construction \textbf{HVor} and \textbf{PVor} are an example of 3D tesselation. Thus, like for planar networks in 2D such as the \textbf{HCN} and \textbf{PCN}, low values of Cheeger mixing are expected. On the contrary, the \textbf{RRN} offers an extreme example of expected high values of Cheeger mixing (Figure \ref{fig:big_graphs}). These three models constitute null models to place the \textbf{LNCN}. We find that the \textbf{LNCN} have similar values of Cheeger as the \textbf{HVor} and \textbf{PVor}. Thus, in the \textbf{LNCN}, the highly connected nodes are also topologically close.

The \textbf{LNCN} is more heterogeneous than both Voronoi tesselations and Random regular network. Because the \textbf{RRN} is random, it is not expected to have any structure, it is thus a good instance of null model. Similarly the \textbf{HVor} was built as a Voronoi tesselation of a 3D regular grid so we also expect it to be an example of very homogeneous network. As for the \textbf{PCN}, because it was built as a Voronoi tesselation of points sampled into a two-peaks 3D Gaussian distribution, it is expected to be a rather heterogeneous network.
As expected the $<p_{in}>_C$ and $<p_{out}>_C$ measures reveal that \textbf{RRN} and the \textbf{HVor} are very homogeneous: for all times, the $<p_{in}>_C$ and $<p_{out}>_C$ difference between the clusters are small, as measured by $\max_{t}SD_{C}(<p_{in}>_C(t))$ and $\max_{t}SD_{C}(<p_{out}>_C(t))$.  The \textbf{PVor} is more heterogeneous than both the \textbf{RRN} and the \textbf{HVor}. The \textbf{LNCN} is about twice as heterogeneous as the \textbf{PVor}.

We have shown that the \textbf{LNCN} is spatially coherent, and we measured the heterogeneity in the \textbf{LNCN} across time, showing that its level of heterogeneity is significant compared to null models. In the following section we will locate spatially this heterogeneity, specifically at the time scale that is relevant for T cells scanning behavior in order to draw biological conclusion.

\subsection* {The \textbf{LNCN} displays heterogeneities localized in small specific regions, and is homogeneous on most of its total volume, promoting uniform exploration behavior \label{LN_bio}}

\begin{figure*}[!ht]
\centering
\includegraphics[width = 0.5\textwidth]{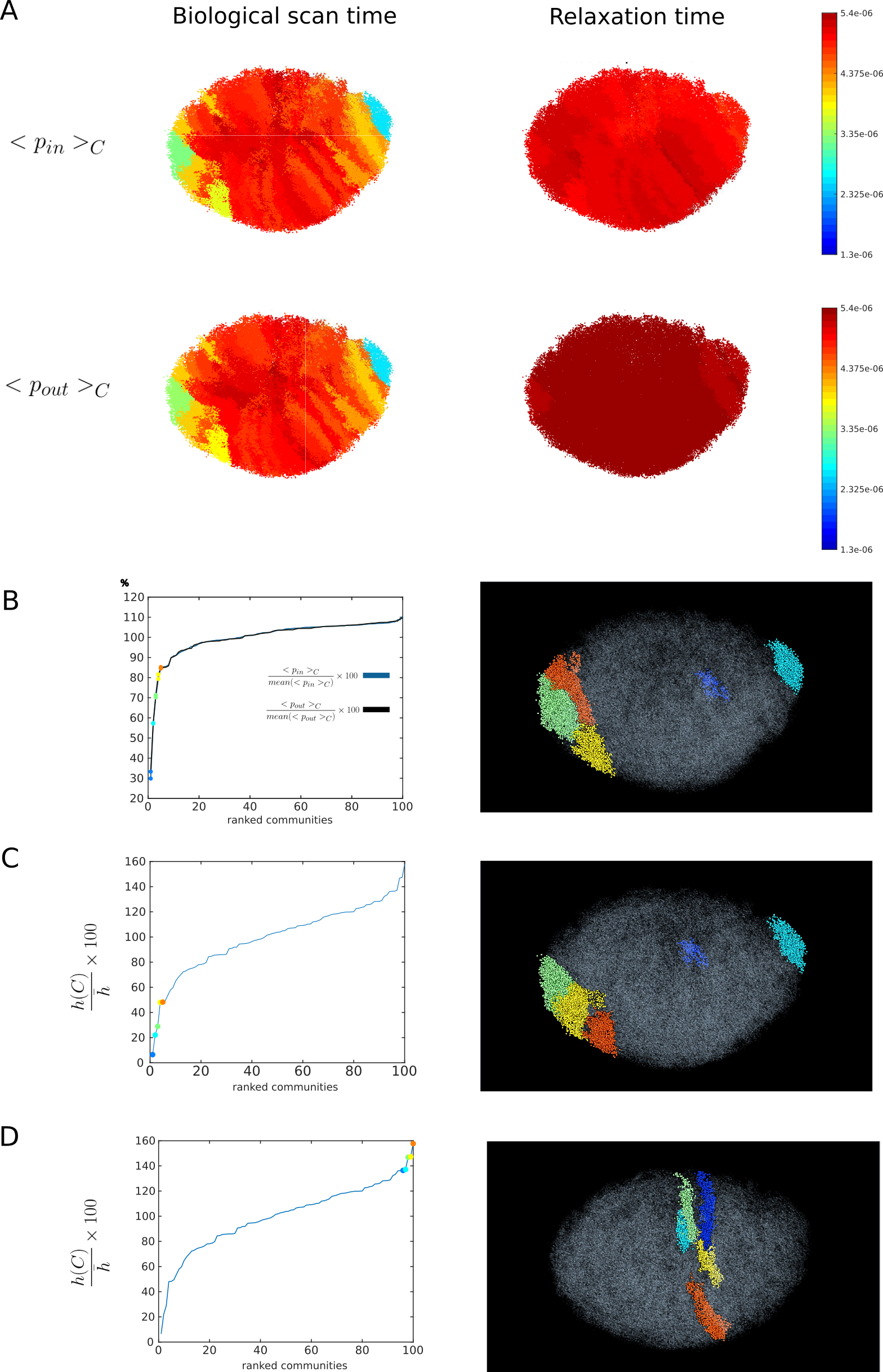}
\caption{A: $<p_{in}>_C$ and $<p_{out}>_C$ for each community is biological scan time t=935. For comparison on the right of $<p_{in}>_C$ and $<p_{out}>_C$ for relaxation time with the same color scale. The variations are higher at biological scan time that at relaxation time. B: On the left panel, $<p_{in}>_C$ and the $<p_{out}>_C$ at biological scan time showed as percentage of the mean value across the 100 communities, shown for each of the 100 communities, ranked from smallest to largest value. The five communities with smallest values of $<p_{in}>_C$ and $<p_{out}>_C$ are shown as colored dots. On the right panel these five communities locations are shown with the same colors. C,D: On the left panels, Cheeger mixing value of each community represented as percentage of the mean value across the 100 communities, shown for the 100 communities, ranked from smallest to largest value. The five communities with smallest (C), and highest highest (D) values are shown as colored dots. On the right panels, these communities locations are shown with the same colors.}
\label{fig:LN_bio}
\end{figure*}

The biological scanning time corresponds to t=935 steps. Indeed, the CD4+ T cells stay about 12h in the lymph node. With an estimation that the naive T-cells move with an average speed of $13\mu m.min^{-1}$ \cite{miller2002two} and that the average length of an edge is $10\mu m$ \cite{FRC_small_world}. We can make the rough estimation that one time step represents 0.77 min. Thus, the 12h exploration time corresponds to t=935 step time. We show in Figure \ref{fig:LN_bio} the $<p_{in}>_C$ and the $<p_{out}>_C$ for this time step. For comparison we show with the same color bar these values for the same communities at relaxation time as already showed in Figure \ref{fig:LN}.
At time t=935, for each cluster, $<p_{in}>_C$ is more similar to $<p_{out}>_C$ than at $t=\tau$. The level of heterogeneity, measured by the standard deviation of  $<p_{in}>_C$ and $<p_{out}>_C$ across all the communities, is higher than for $t=\tau$ ( $<p_{in}>_{C}(t=935)=4.8\times 10^{-7} , <p_{out}>_{C}(t=935)=4.7\times 10^{-7} $ whereas $<p_{in}>_{C}(t=\tau)=1.1\times 10^{-7},<p_{out}>_{C}(t=\tau)=0.25\times 10^{-7} $)

Heterogeneities are localized in restricted regions on the long axis extremities and near the medulla. Figure \ref{fig:LN_bio} shows the values of $<p_{in}>_C(t=935)$ and $<p_{out}>_C(t=935)$ ranked from smallest to largest across all 100 communities.
We located the 5 communities with smallest $<p_{in}>_C(t=935)$ and $<p_{out}>_C(t=935)$. They are located at the extremities of the long axis and near the medulla (dark blue). 

We also computed the 5 communities with respectively the smallest and largest Cheeger mixing (Figure \ref{fig:LN_bio}. Among the 5 communities with smallest Cheeger mixing index, 4 of them are also part of the 5 communities with smallest $<p_{in}>_C(t=935)$ and $<p_{out}>_C(t=935)$. The 5 communities with largest Cheeger mixing index are communities which shapes are slices cutting the longest axis at its center.  

In summary, the \textbf{LNCN} promotes spatially continuous and overall uniform random walk. The low Cheeger mixing shows that similarly to a tesselation, in the \textbf{LNCN} the regions that are highly connected by short random walk paths are also close in the network in the sense of topological distance. Compared to the null models that we explored, the level of heterogeneity is significant. We distinguish a small set of communities which have lower $<p_{in}>_C$ and  $<p_{out}>_C$ as well as small Cheeger mixing values.  They are restricted to extremities along the long axis and the medulla. Since the entry and exit probabilities are low they are not drawing T cells faster than other regions and thus are not particularly privileged location for the DCs to sit to be found by the T cells. The rest of the network (most of the volume) is homogeneous, thus overall the network promotes uniform exploration. We also note that the communities, which are sets of nodes that are highly connected when integrating on all paths (diffusion), have anisotropic shapes. There are forming slices along the x axis. Thus it is easier to diffuse in the y and z direction than in the x direction.

\section*{Discussion}
In this paper we characterize a large biological network, the \textbf{LNCN}, which among other functions, serves as a substrate guiding T cells migration in their exploration of the lymph node. We addressed the question of how does the connectivity of the network alone influences the exploration behavior modeled as a random walk on network. We show that the \textbf{LNCN} is as spatially coherent as a 3D tesselation of the same size. Moreover, despite its quasi-regularity, we find some random walk related heterogeneities. This is achieved by applying a same workflow of diffusion based community detection and analysis on the \textbf{LNCN} and a series of generated toy networks.  

The random walk can be tracked analytically through the probability of presence field computed from the transition matrix, using an approximation for large network. This provides a first highlight on the spatial organization of the network as shown in the online visualization platform. Then, by defining diffusion communities and measuring the Cheeger mixing in these communities, we show that the \textbf{LNCN} is spatially coherent, which means that the nodes best connected through random walk paths are also close in terms of number of edges apart.  Furthermore, the diffusion communities display different values of mean entry and exit probabilities $<p_{in}>_C$ and $<p_{out}>_C$. The \textbf{LNCN} displays heterogeneities localized in small specific regions: there are regions at the long axis extremities and near the medulla with significantly different diffusion properties. But elsewhere, the \textbf{LNCN} is homogeneous on most of its total volume, promoting uniform exploration behavior.

We have restricted the analysis to the topology of the network, and we have not included the geometric information: edges lengths and angles. For a more complete view of the network, including some assumptions on the preferential angles can easily be described by a weighted network, so that edges which are for instance most aligned with the one the walker comes from have a higher probability to be chosen. Taking into account the edges lengths, and subsequent traveling times, requires to address continuous time random walk to account for the fact that some edges take more time than others to be crossed, which can also be formulated as: the waiting time between two jumps will be long if the second jump involves crossing a long edge.

Our conclusions about the exploration behavior of the T cells also suffers several limitations. The measures were done on one sample of lymph node. Including more samples would increase the robustness of the conclusions. Ideally, staining of new lymphatic conduits networks would be done simultaneously with the staining of the entry and exit locations of the lymph nodes, respectively the HEVs (high endothelial venules) which are part of the blood vascular network, and the efferent lymphatic vessels at the medulla. This way, one would be able to include inlet and outlet in the random walk on network analysis. Another point that would need clarification is the radical mismatch between the FRC network and the conduits network characteristics. The FRC network, in which nodes and edges are respectively the nuclei and cellular protrusions of the cells that ensheath the conduits, based on a slice of lymph node, shows small world properties and a degree distribution far less regular than the conduits network. We assume, like Kelch et al., that the T cells guiding is more accurately described taking the conduits as substrate. However, if experimental data shows otherwise, our workflow can easily be applied on the FRC network when the topological data of whole lymph node FRC network becomes available.

The question we addressed is to test how the network connectivity influences the exploration behavior of the T cells. This question was motivated by a more general question which is: is the network optimized for its function? In this respect, our finding that the \textbf{LNCN} promotes overall uniform exploration is compatible with an optimal search strategy. Indeed, if on the contrary case, some regions were hubs that draw the T cells in faster than other regions, assuming that dendritic cells sit in these strategic places to be found, T cells might endure traffic jam in these regions. Thus we covered the relation between the network structure and its T cells guiding function. However, the conduits is also a piping system in which lymph flows, conveying crucial immune system molecules such as antigens, inflammatory soluble mediators and cytokines across the lymph node. Thus its optimality could also be analyzed from a hydrodynamics point of view such as modeled in the cortex capillary network \cite{goirand2021networkNC, goirand2021network}. 

Altogether, this study provides a general pipeline for the analysis of large networks at low computational cost and paves the way for the characterization of heterogeneity within biological tissue.  

\section*{Supporting information}

\paragraph*{S1 Appendix.}
\label{S1_Appendix}
{\bf Detailed spectral decomposition of the transition matrix} 

$T$ is non symmetric so its spectral decomposition is not guaranteed. Let $T_{s}=D^{1/2}TD^{-1/2}$ the symmetrized transition matrix. $T_s$ can be composed as $T_s = V \Lambda V^{T}$, where $\Lambda$ is a diagonal matrix with the eigenvalues of $T_s$, which as the same as the eigenvalues of $T$  and $V$ is an orthogonal matrix ($VV^{T}=Id$) which columns are the eigenvectors of $T_s$. Then,
\begin{equation}\label{T_exp}
\begin{aligned}
   T&=D^{-1/2}T_{s} D^{1/2}\\
   &=D^{-1/2}V\Lambda V^{T} D^{1/2}\\
   &=\Psi\Lambda\Phi^{T} 
\end{aligned}
\end{equation}
with $\Psi=D^{-1/2}V$ and $\Phi=D^{1/2}V$

\paragraph*{S2 Appendix.}
\label{S2_Appendix}
{\bf Numerical approximation of the spectral decomposition for the large networks} 

The computation of the full spectral decomposition of a symmetric matrix $A$ of size $N \times N$ requires $O(N^3)$ floating point operations and $O(N^2)$ coefficients to store, which becomes intractable even for medium-sized $N$.
In particular, this algorithm fails for the \textbf{LNCN} dataset where $N = 200,000$.

However approximating $T^t$ by $\widehat{T}_k^t = V_k \Lambda_k^t V_k^T$ only requires the computation of the truncated decomposition $V_k \Lambda_k V_k^T$ where $V_k$ is of size $N \times k$ and contains the eigenvectors associated to the $k$ largest eigenvalues and $\Lambda_k = \textrm{diag}(\lambda_0, \ldots, \lambda_k)$ is a diagonal matrix of size $k \times k$. 
This observation can be leverage to implement spectral decomposition algorithm that are fast provided that $k \ll N$.

We used the \emph{eigsh} function from the \emph{scipy} library which implements a Lanczos method \cite{lehoucq1998arpack,golub1996matrix}. 
This method is iterative and mainly needs the computation of $k$ matrix-vector products and an orthogonalization step of complexity $O(k^2 N)$.
For general matrices, the cost of $k$ matrix-vector products is $O(k N^2)$, but for sparse matrices, this complexity reduces to $O(k L N)$ where $L$ is the average number of non-zero coefficients in each row. The total complexity of the Lanczos method for sparse matrices is therefore bounded by $O(k^2 N + k L N)$.
For the \textbf{LNCN} dataset, the matrix $T_s$ contains roughly $L=3$ non-zeros per row, leading to a very efficient diagonalization procedure.

\paragraph*{S3 Appendix.}
\label{S3_Appendix}
{\bf Estimation of the error of the approximation of the spectral decomposition of the transition matrix} 

In this section we describe the computation of the approximation error between $T^t$ to $\widehat{T}_k^t$ as defined in equation (9) in the main part of the paper.

Let $A$ be a matrix of size $M \times N$, its spectral norm $\| A \|_{2,2}$ is given by 
\begin{equation}
  \| A \|_{2,2} = \sup_{x \in \mathbb{R}^N, \| x\|_2 = 1} \| A x \|_2.
\end{equation}
Importantly, the spectral norm of a matrix is equal to its largest singular value i.e. $\| A \|_{2,2} = \sigma_1(A)$.

This norm induces the spectral distance $\| A - B \|_{2,2}$ between two matrices of size $M \times N$ and measures the maximal error that can be made between the two vectors $Ax$ and $Bx$ with $x \in \mathbb{R}^N$.

Let $\Delta = T^t - \widehat{T}^t_k$. The computation of $\| \Delta\|_{2,2}$ requires the computation of  $\sigma_1(A) = \sqrt{\lambda_1(\Delta^T \Delta)}$, with $\lambda_1(\Delta^T \Delta)$ refering to the largest eigenvalue of $\Delta^T \Delta$.

Large graphs prevents the matrix $\Delta^T \Delta$ to be stored in memory, we therefore need an indirect way of computing its largest eigenvalue.
The power iteration method \cite[Section 8.2.1 p406]{golub1996matrix} computes the largest eigenvalue of a given matrix using only its matrix-vector products, which makes it suitable for large scale graphs.

We implemented the power method on $\Delta^T \Delta$ through the two the matrix-vector products with $\Delta^T$ and $\Delta$.
This products are made of:
\begin{itemize}
\item $t$ matrix-vector products with the sparse matrix $T$ and $T^T$.
 For a matrix of $LN$ non-zero coefficients, this step costs $O(t L N)$ operations.
\item matrix-vector products with $\widehat{T}^t_k = \Psi \Lambda_k^t \Phi^T$ and $(\widehat{T}^t_k)^T =  \Phi \Lambda_k^t \Psi^T$ which both cost $O(kN)$ operations.
\end{itemize}
For the \textbf{LNCN} dataset, the matrix $T$ contains roughly $L = 3$ non-zeros per row, leading to a fast computation of $\| \Delta \|_{2,2}$.

We define the relative error as $\frac{\| \Delta \|_{2,2}}{\| T^{t} \|_{2,2}}$, with $\| T^{t} \|_{2,2}$ also computed with power iteration method.

\paragraph*{S1 Fig.}
\label{S1_Fig}
{\bf Relative error estimation of the approximated transition matrix for the \textbf{LNCN}.} Relative spectral norm between the exact transition matrix and the approximated one with truncation at k=2000 first eigenvalues, decaying with time steps. Insert: log scale representation.

\begin{figure}[htbp]
  \centering
  \includegraphics[width = 0.9\textwidth]{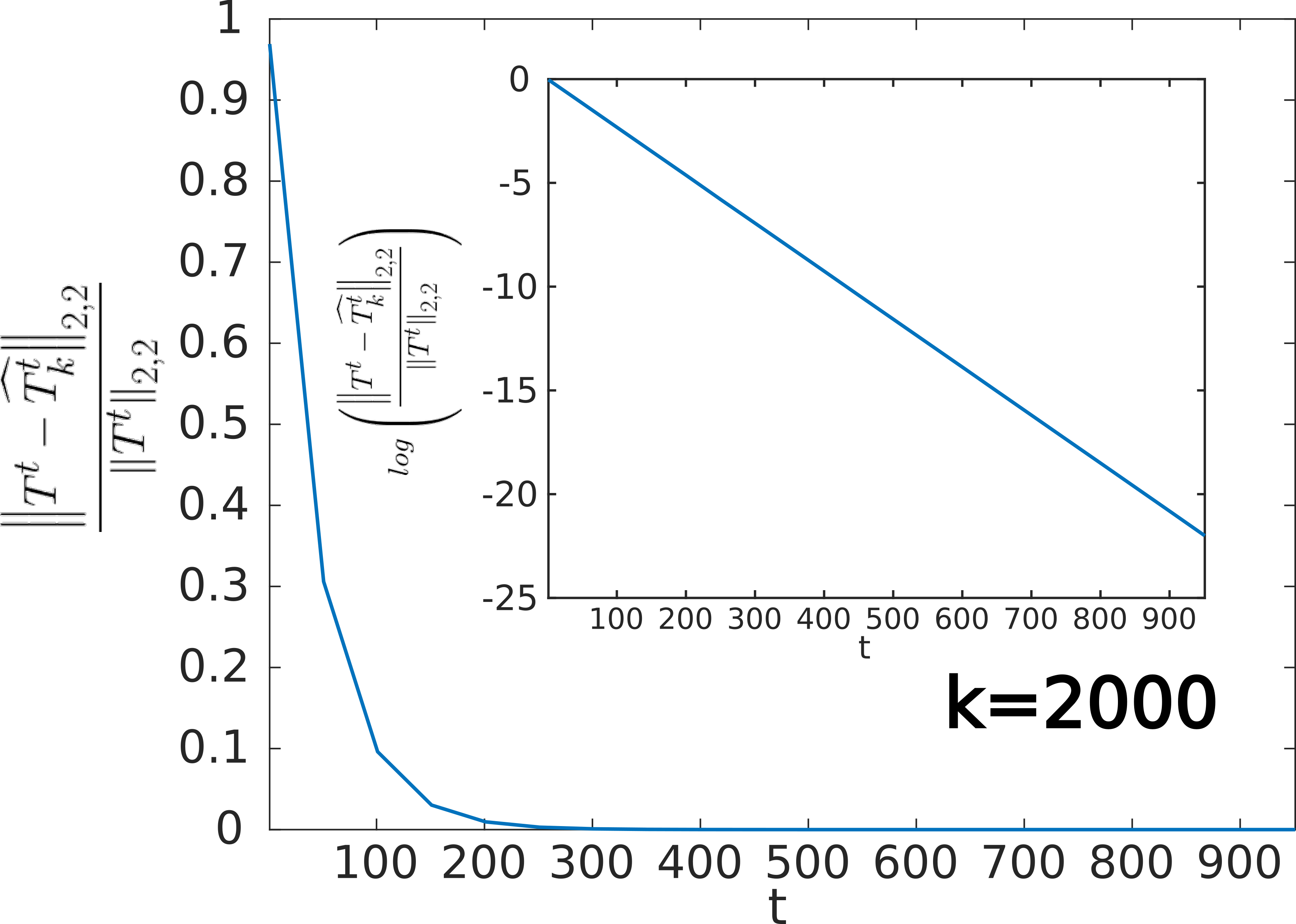}
  \label{fig:LN_error}
\end{figure}

\paragraph*{S4 Appendix.}
\label{S4_Appendix}
{\bf Relaxation time calculation} 

The relaxation time $\tau$ is defined as the time for which

$|p(i,i|t)-p(j,t=\infty)|\leq e^{-1}$

According to  https://www.stat.berkeley.edu/~aldous/RWG/Chap4.pdf, 

$|p(j,i|t)-p(j,t=\infty)|\leq\lambda_{1}^{t}$

Thus the condition 

$|p(j,i|t)-p(j,t=\infty)|=e^{-\frac{t}{\tau}}=\lambda_{1}^{t}$

gives

\begin{equation}\label{relax}
\begin{aligned}
    e^{-\frac{t}{\tau}}&=\lambda_{1}^{t}\\
    e^{-\frac{t}{\tau}}&=e^{t\log{\lambda_{1}}}\\
    -\frac{t}{\tau}&=t\log{\lambda_{1}}\\
    -\frac{t}{\tau}&=t\log(1-(1-\lambda_{1}))\\
    -\frac{t}{\tau}&\sim -t(1-\lambda_{1})\\
    \tau&\sim \frac{1}{1-\lambda}
\end{aligned}   
\end{equation}

\paragraph*{S5 Appendix.}
\label{S5_Appendix}
{\bf City street network generative model -\textbf{HCN} \& \textbf{PCN}} 

\textbf{HCN} and \textbf{PCN} are two instances of generative model for city streets morphogenesis \cite{pousse2020caracterisation}. 
The networks are constructed by iterative subdivisions of an initial rectangle. The streets segments are the rectangle boundaries. At each generation, the rectangle with the highest score $U_{rec}$ is divided. $U_{rec}$ reads:
$ U_{rec}=\left[\sum\limits_{i=1}^{N_{i}}\frac{1}{d(C_{rec},i)^\alpha}\right]\times [L_{rec}]^{\beta}$ where $d(C_{rec},i)$ is the distance from the center of mass of the cell of interest to the node $i$ and $L_{rec}$ is the length of the cell.

The first term favours further subdivision of the high density zones while the second term favours the divisions in the low density zones. Their respective weights are tuned by the choice of $\alpha$ and $\beta$. When $\alpha > \beta$ the morphogenetic process is leading to cities with dense poles. On the other hand when $\alpha < \beta$ the morphogenetic process is leading to homogeneous streets density.

The \textbf{HCN} (Homogeneous City Network) and \textbf{PCN} (Polar City Network) were generated respectively with $\alpha < \beta$ and $\alpha > \beta$. 

Both networks contain 2050 nodes and 3073 edges.

\paragraph*{S6 Appendix.}
\label{S6_Appendix}
{\bf Interactive visualization}

\underline{Description of the scripts}
The code is developed in Python. There are 2 files :  "graph\_object.py"  and "main.py". The script "graph\_object.py"  defines a class object for a network. All the functions needed to either load a graph and compute analysis are defined in it.
The "main.py" script aims to both compute the network analysis and create figures. At the end of the script, the interactive application is deployed to visualize all the analysis. We used the module Plotly for the figures and Dash for the app.

\underline{Centralities}

\begin{itemize}
    \item Betweenness centrality: 
    The betweenness centrality value of a node $u$ is the number of shortest paths between two nodes that passes through that given node. 

\begin{equation}
C_{btw}(u)=\sum_{s,t\in V}\frac{\sigma(s,t|u)}{\sigma(s,t)}
\end{equation}
where $\sigma(s,t)$ denotes the set of paths connecting $s$ and $t$ and $\sigma(s,t|v)$ is the set of paths connecting $s$ and $t$ and going through node $u$.
The betweenness centrality was calculated using the NetworkX Python library \cite{ref_networkx} networkx.algorithms.centrality.betweenness\_centrality implemented from \cite{ref_betweenness}
    \item Closeness centrality
    The closeness centrality of a node is computed as the inverse of the sum of the length of the shortest path to all the nodes in the network.
\begin{equation}
C_{cls}(u)=\frac{n-1}{\Sigma_{v \in V} d(u,v)}
\end{equation} where $ d(u,v)$ is the length of the shortest path between nodes $u$ and $v$, and $n$ the number of nodes that can reach u.
The closeness centrality was calculated using the NetworkX Python library \cite{ref_networkx} networkx.algorithms.centrality.closeness\_centrality implemented from \cite{ref_closeness}.

    \item Maximal remoteness centrality
We define the maximal remoteness centrality as  $C_{maxrm}(u) = \max\limits_{v} d(u,v) $  with 
$d(u,v)$ the length of the shortest path between $u$ and $v$.
This is not a classical measure but it captured well the two high density zones in the polar model.
    \item Eigenvector centrality
The eigenvector centrality \cite{ref_eigen_centrality} is the score obtained by computing the eigenvector of the adjacency matrix with the highest eigenvalue $Ax = \lambda x$ using eigs function in Matlab. It is interpreted as a centrality measure in which a node can have a high centrality value if it is connected to few other nodes but those nodes have a high centrality value.
    \item GMFPT
We calculate this value following:
\begin{equation}
    GMFPT(x)=\frac{1}{p(x,t=\infty)}\sum^{\infty}_{n=0}p(x,t=n|x)-p(x,t=\infty)
\end{equation}
The numerical calculation is truncated following
\begin{equation}
    GMFPT(x)\approx \frac{1}{p(x,t=\infty)}\sum^{n_{cut}}_{n=0}p(x,t=n|x)-p(x,t=\infty)
\end{equation}
where $\frac{|p(x,t=n_{cut}|x)-p(x,t=\infty)|}{\frac{1}{N}\sum_{x}p(x,t=\infty)}<0.01$
    \item Normalization
All the centrality measures are then normalized following $c_{norm}(x)=\frac{c(x)-min(c)}{max(c)-min(c)}$

\end{itemize}

\paragraph*{S7 Appendix.}
\label{S7_Appendix}
{\bf Computation of Node2vec communities}

The Node2vec algorithm \cite{grover2016node2vec} provides an embedding of the nodes of the network that preserves the neighbourhoods of the nodes, the neighborhoods being sampled by simulated biased random walks. The bias is defined by setting p and q, respectively the return parameter and the in-out parameter. At each step, the probability of going back to the previous node is $\frac{1}{p}$ and the probability of going to any other neighbouring node is $\frac{1}{q}$ provided that they are not also immediate neighbours of the previous node. Random walks are simulated and samples of $k$ successively occupied nodes are recorded. Theses sampling define the neighbourhoods of nodes. A stochastic descent is run to embed the nodes in a $d$ dimensions space while preserving as best as possible the neighbourhood relationships. We computed the embeddings using the PecanPy Python implementation \cite{liu2021pecanpy}. We explored the parameters to optimize the embeddings so to get clusters as similar as possible to ours with the procedure described below.  The clusters obtained with the set of parameters $\{d=10, q=0.1, p=5, k=50\}$, optimized for the \textbf{PCN}, are shown in main Figure 3.M-N .

\paragraph*{S8 Appendix.}
\label{S8_Appendix}
{\bf Node2vec parameters optimization}

The parameters to optimize are d (the dimension of the embedding), p (the return parameter), q (the in-out parameter), and k (the length of the walk).
The optimization was done with a combination of two criteria that evaluate the 4-clusters. The first one is the Cheeger mixing $h(C)$ and the second one is the mean Jaccard distances between their clusters and ours, after optimizing the pairing of clusters. We tested all the combinations of parameters in the sampling $p\in\{0.05,0.1,0.5,1,5,10,20\}$, $q\in\{$0.05,0.1,0.5,1,5,10,20\}, $k\in\{3,5,10,20,50\}$ and $d\in\{2,3,5,10,50,100,200\}$. We compared the two populations $p<min(q,1)$ and $p>max(q,1)$. The scores $s(\{C\})=\frac{\bar h+<J(\{C\})>}{2}$ (Figure S2.A) were smaller for the population $p>max(q,1)$, which corresponds to the no-return configurations. We kept only the combination of parameters with $p>max(q,1)$. Among these combinations, we computed the correlations between the combined score s(\{C\}) and each of the other parameters p,q,k and d. There was a weak anti-correlation with $k$ (R=-0.55) so we set $k=max(\{k\}=50)$. Then, we ranked the remaining combinations by ascending scores and we plotted the corresponding scores. The curve shows a steep slope then a plateau (Figure S2.B). We picked the 10 best combinations with lowest score s(\{C\}). We note that they all present rather compact clusters in space, but they all show a band-like structure. Thus, the embeddings by Node2vec lead to different communities than the diffusion communities.

\paragraph*{S2 Fig.}
\label{S2_Fig}
{\bf Node2vec parameters optimization and communities for \textbf{PCN}.} A)Comparison of no return and with return sets of parameters. No return:$p>max(q,1)$. With return:$p<min(q,1)$. The score $S(\{C\})$ to be minimized combines the mixing index and the Jaccard distance to the clusters make on the diffusion space at t=500 shown previously. C)4-means clustering done on Node2vec embedding with parameters $\{d=10, q=0.1, p=5, k=50\}$, which gives the lowest score.
\begin{figure}[!htbp]
  \centering
  \includegraphics[width = 0.5\textwidth]{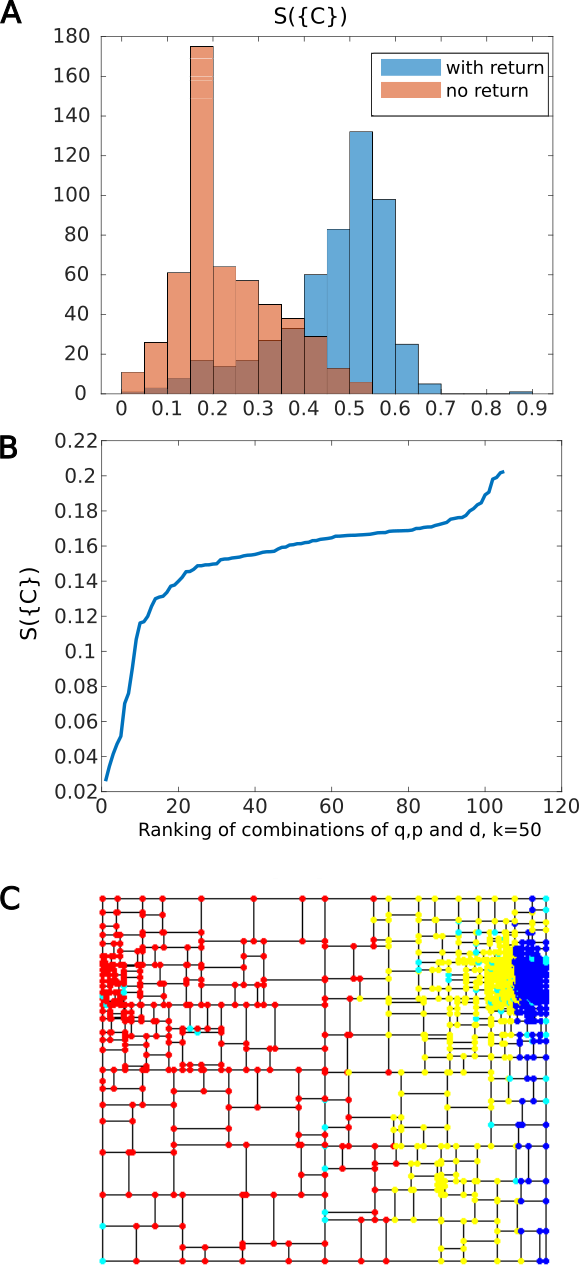}
  \captionsetup{labelformat=empty}
  \caption{\bf S2 Fig}
  \label{fig:node2vec_suppl}
\end{figure}

\paragraph*{S9 Appendix.}
\label{S9_Appendix}
{\bf Features of $<p_{in}>_C$ and $<p_{out}>_C$ }
\begin{itemize}
    \item $<p_{in}>_{C}(t)$=$<p_{out}>_{C}(t)$ for regular graphs
    
    \begin{equation}
    <p_{in}>_{C}(t)=\frac{1}{|C||\Bar{C}|}\sum_{i \in C}\sum_{j \in \Bar{C}}\sum_{k}\psi_{jk}\lambda_{k}^{t}\phi_{ik}
\end{equation}

\begin{equation}
     <p_{out}>_{C}(t)=\frac{1}{|C||\Bar{C}|}\sum_{i \in C}\sum_{j \in \Bar{C}}\sum_{k}\psi_{ik}\lambda_{k}^{t}\phi_{jk}
\end{equation}

Given that $\forall i, \phi_{ik}=\frac{\psi{ik}}{d_i}$,

\begin{equation}
     <p_{in}>_{C}(t)=\frac{1}{|C||\Bar{C}|}\sum_{i \in C}\sum_{j \in \Bar{C}}\sum_{k}\psi_{jk}\lambda_{k}^{t}\frac{\psi_{ik}}{d_i}
\end{equation}

and 

\begin{equation}
<p_{out}>_{C}(t)=\frac{1}{|C||\Bar{C}|}\sum_{i \in C}\sum_{j \in \Bar{C}}\sum_{k}\psi_{ik}\lambda_{k}^{t}\frac{\psi_{jk}}{d_j}
\end{equation}

Thus if the graph is regular $\forall i,j$
 $d_i=d_j=d$

Then $<p_{in}>_{C}(t)=<p_{out}>_{C}(t)$

\item $<p_{in}>_{C}(t)$ and $<p_{out}>_{C}(t)$ converge to respectively the relative volume of C and $\Bar{C}$ when $t\longrightarrow \infty$

Given that $1=|\lambda_0|>|\lambda_1|>...>|\lambda_N|>0$

$
<p_{in}>_{C}(t)\underset{t\to +\infty}{\longrightarrow}\frac{1}{|C||\Bar{C}|}\sum_{i \in C}\phi_{i0}\sum_{j \in \Bar{C}}\psi_{j0}$

Which can be re-writen

$
\frac{1}{|C||\Bar{C}|}\sum_{i \in C}\phi_{i0}\sum_{j \in \Bar{C}}\psi_{j0}=\frac{1}{|C||\Bar{C}|}\frac{1}{d_{tot}}\sum_{i \in C}d_i\sum_{i \in |\Bar{C}|}1=\frac{1}{|C|}\frac{\sum_{i \in C}d_i}{d_{tot}}
$

Thus, $<p_{in}>_{C}(t)\underset{t\to +\infty}{\longrightarrow}\frac{1}{|C|}\frac{\sum_{i \in C}d_i}{d_{tot}}$

Similarly $<p_{out}>_{C}(t)\underset{t\to +\infty}{\longrightarrow}\frac{1}{|\Bar{C}|}\frac{\sum_{i \in \Bar{C}}d_i}{d_{tot}}$

\end{itemize}

\paragraph*{S3 Fig.}
\label{S3_Fig}
{\bf diffusion communities $<p_{in}>_C$ and $<p_{out}>_C$ for toy networks}  A:geometric network B:RRN C:ER networks
\begin{figure}[!htbp]
  \centering
  \includegraphics[width = 0.6\textwidth]{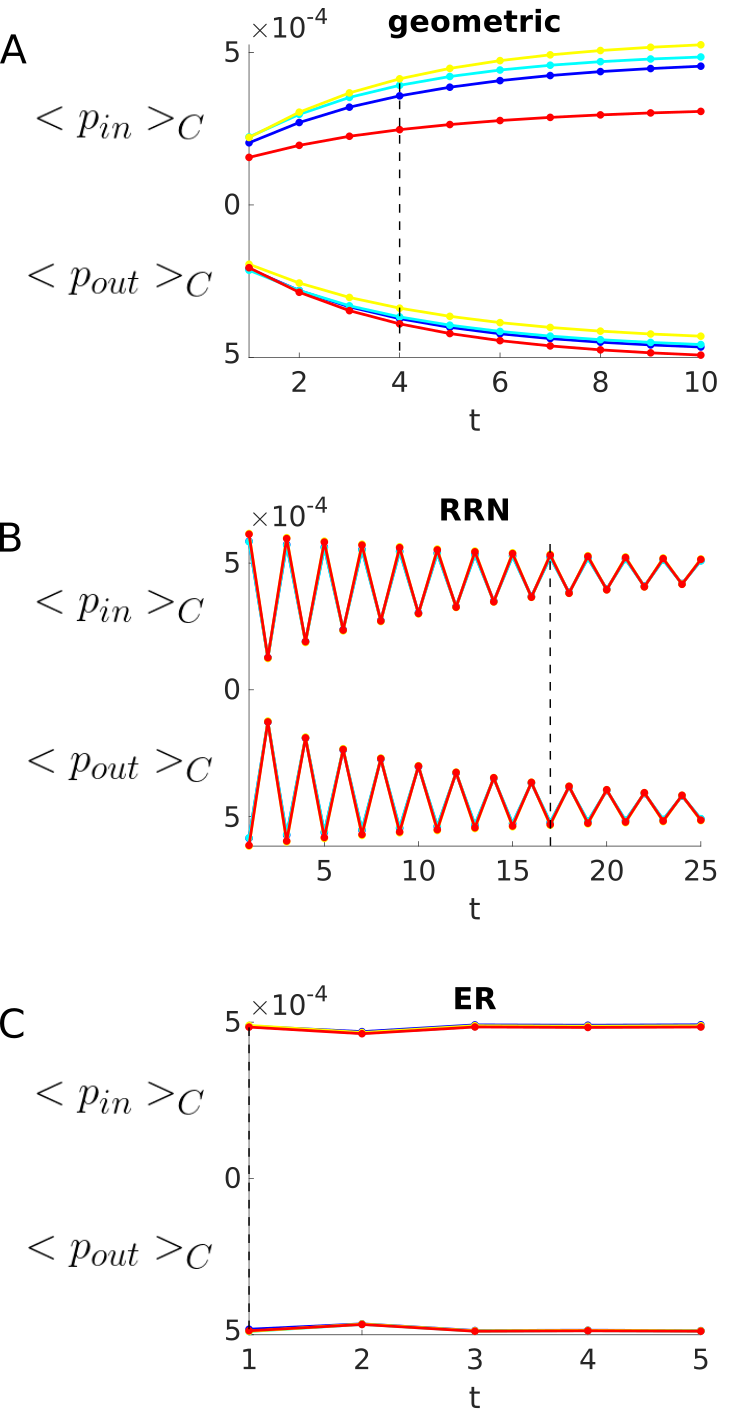}
  \captionsetup{labelformat=empty}
  \caption{\bf S3 Fig}
  \label{fig:pin_pout_toys}
\end{figure}

\paragraph*{S4 Fig.}
\label{S4_Fig}
{\bf Variation of $\bar h$ with respect to the number of clusters}  For each of the toy models of small size: \textbf{HCN}, \textbf{PCN}, \textbf{Geometric}, \textbf{RRN} and \textbf{ER}, we made k clusters using k-means in the diffusion space with k varying from 2 to 10. For each value of k, we computed the mean value of the Cheeger mixing index over all k communities. Except from the case of the \textbf{ER}, the the mean Cheeger mixing index consistently increases with the number of clusters, as the clusters become smaller. This effect is analogous to the increase of surface to volume ratio as the size decreases in the physical space. Here the surface is the nodes at the interface between different communities. This shows that the mixing index is very number of clusters dependent and thus justifies to stick with one given number of communities for all networks of the same size to be compared. 
\begin{figure}[!htbp]
  \centering
  \includegraphics[width = 0.8\textwidth]{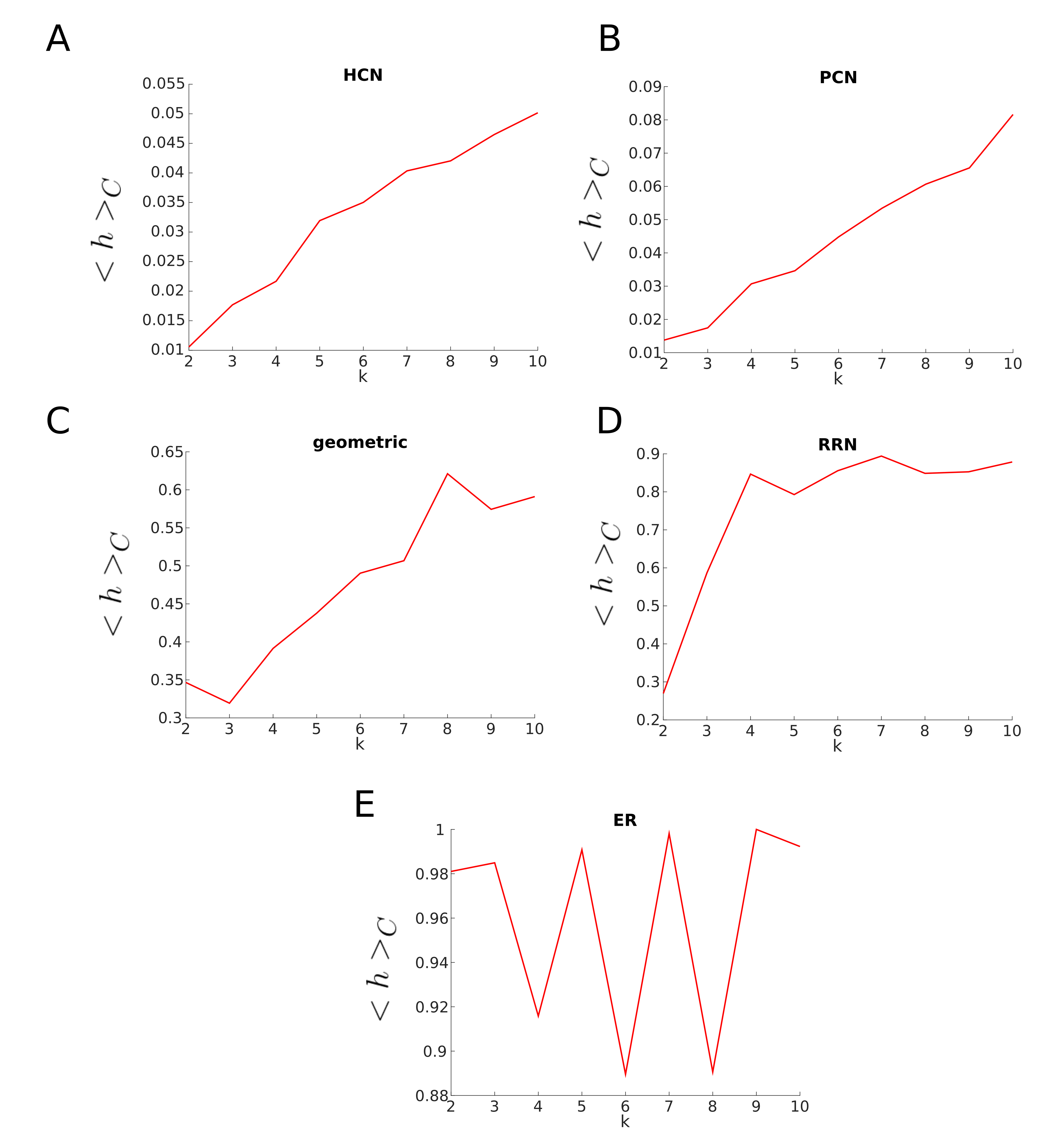}
  \captionsetup{labelformat=empty}
  \caption{\bf S4 Fig}
  \label{fig:mean_h_different_k}
\end{figure}

\section*{Acknowledgments}
We thank Marc Bajénoff for initial discussions on the biological aspects of T cells migration in the lymph node. We thank Inken Kelch, Gib Bogle and Rod Dunbar for providing the lymph node conduits data, and fruitful discussions. We thank Romain Pousse and Stéphane Douady for prodiving the toy networks simulated from their city morphogenesis model, and fruitful discussions. We thank Ana\"is Baudot, Anthony Baptista and Alain Barrat for their critical reading of the manuscript and helpful discussions. We thank Nicolas Levernier, Jean-François Rupprecht and Tanguy Fardet for helpful discussions. We thank Guillaume Gay from the Multi-Engineering Platform of the Turing Center for Living Systems, Marseille, France, for his valuable support on the development of the online interactive visualization tool.
S.S., M.S. and P.V. were funded by the "Investissements d'Avenir" French Government program managed by the French National Research Agency (ANR-16-CONV-0001) and from Excellence Initiative of Aix-Marseille University - A*MIDEX. P.E. was funded by CNRS.


%
%
%


\end{document}